\documentclass[12pt]{article}
\usepackage{epsfig}
\usepackage{amssymb}
\usepackage{amsmath,amsfonts,amssymb,graphicx}
\newcommand {\be}{\begin{equation}}
\newcommand {\ee}{\end {equation}}
\newcommand{\beq}{\begin{eqnarray}}
\newcommand{\eeq}{\end{eqnarray}}

\begin{document}
\vspace{-2cm}

\title{Estimate of CP Violation for the LBNE Project and $\delta_{CP}$}
\author{Leonard S. Kisslinger\\
Department of Physics, Carnegie Mellon University, Pittsburgh, PA 15213}
\maketitle
\noindent
PACS Indices:11.30.Er,14.60.Lm,13.15.+g

\vspace{0.25 in}
\begin{abstract}

   Measurements of CP violation (CPV) and the basic $\delta_{CP}$ parameter
are the goals of the LBNE Project, which is being planned. Using the
expected energy and baseline parameters for the LBNE Project, CPV and the
dependence of CPV on $\delta_{CP}$ are estimated, to help in the planning of
this project.
\end{abstract}
\vspace {0.25 in}

\section{Introduction}

  For several decades there have been many experimental and theoretical
studies of CP violation (CPV). Recently we have studied time reversal
violation (TRV)\cite{hjk11} and CPV\cite{khj11} for neutrino oscillations
in matter, using parameters of current neutrino oscillation experiments,
MiniBooNE\cite{mini}, JHF-Kamioka \cite{jhf}, MINOS\cite{minos}, and
CHOOZ\cite{chooz}. See Refs.\cite{hjk11},\cite{khj11} for details and 
references to earlier work on T and CP. When neutrino beams move through 
matter the CPT theorem is not valid (see. e.g., Ref\cite{jo04}); therefore 
TRV and CPV are not directly related. There have been many other studies 
of CP asymmetries in weak decays: see Ref\cite{blnp11} for a recent study
of $\bar{B}$ radiative decay with references to earlier work on CP violation
in various weak decays. Also, in addition to the baseline, energy, and
the matter density, there are a number of parameters associated with neutrino
oscillations, as will be seen in our discussion of CP violation below. There
have been a number of studies of these parameters; see, e.g., 
Refs\cite{bmw02,ajlos04}, which contain references to earlier related 
publications. 

  In our present work we study CPV using the baseline and
energies expected for the future LBNE Project\cite{LBNE}. See 
Ref\cite{LBNE07} for a detailed report on the project, and the recent 
discussion of the LBNE parameters\cite{LBNE11}, with a baseline L=1300 km
and energies in the range E=0.5-12 GeV. We use the expected baseline of
1300 km and five possible energies. There have also been 
a number of studies of matter effects for for the LBNE Project\cite{LBNE07}. 
See Davoudiasl $et \; al$\cite{dlm11}and Gonzalez-Garcia $et \; al$\cite{gms11} 
for recent studies of matter effects and various parameters
important for the  baseline L expected in the LBNE Project,
and references to earlier studies for the LBNE Project. 
 
One of the main objectives of this future project is to measure the 
$\delta_{CP}$ parameter. In the present work we calculate CPV as a function
of $\delta_{CP}$ for parameters being studied for the LBNE Project, to help
in the design of this future experiment. In order to determine $\delta_{CP}$ 
via neutrino oscillations one also needs the angles $\theta_{12},\theta_{23}$,
which are well-known, and the angle $\theta_{13}$, which is being studied
by a number of experiments: T2K\cite{T2K11}, Daya Bay\cite{DayaBay,DayaBay11}
Double Chooz\cite{dc11,dcnov11}, and RENO\cite{RENO}. A very recent result 
from the Daya Bay project\cite{DB3-7-12}, which is consistent within errors
of the recent RENO\cite{RENO} result, finds that $s_{13} \simeq 0.15$, 
which we use in the present study.

 As pointed out above, an essential aspect of the determination of CPV, as 
well as TRV and CPTV is the interaction of neutrinos with matter as they 
travel along the baseline. See, e.g., Refs\cite{as97,bgg97,kty02,ahlo01}. 
Since this was discussed in detail in Ref\cite{khj11}, some details are not 
given in the present work.
In the present work we use the notation and formalism of 
Jacobson and Ohlsson\cite{jo04}, who studied possible matter effects for 
CPT violation. 
 
  CP violation in the $a-b$ sector is given by the transition probability,
denoted by $\mathcal{P}(\nu_a \rightarrow \nu_b)$, for a neutrino of flavor
$a$ to convert to a neutrino of flavor $b$; and similarly for antineutrinos
$\bar{\nu}_a,\bar{\nu}_b$.
   The CPV (note that the C operator changes a
particle to its antiparticle) is defined as
\beq
   \Delta\mathcal{P}^{CP}_{ab}&=& \mathcal{P}(\nu_a \rightarrow \nu_b)
-\mathcal{P}(\bar{\nu}_a \rightarrow \bar{\nu}_b) \; .
\eeq

In our present work we study $\mathcal{P}(\nu_\mu \rightarrow \nu_e)$ and
$\mathcal{P}(\bar{\nu}_\mu \rightarrow \bar{\nu}_e)$, since the neutrino 
beams at LBNE are muon or anti-muon neutrinos.

As discussed above, there are four angles in the matrix relating a neutrino
with flavor to neutrinos with mass, the basis for neutrino 
oscillations. The two angles under current study are $\delta_{CP}$ and 
$\theta_{13}$. In order for the LBNE proposed project to be successful in 
determining $\delta_{CP}$, $\theta_{13}$ must be known. One might expect that
the experiments Daya Bay, RENO, and Double Chooz could not achieve
their goal of determining $\theta_{13}$ to an accuracy of 1\% without knowing
$\delta_{CP}$. As is discussed below, fortunately, this is not
the case.
\newpage

\section{CP Violation $\Delta \mathcal{P}^{CP}_{\mu e}$}

\indent

   We use the time evolution matrix in flavor space to derive CPV.
The neutrino state at time = $t$ is obtained from the state at time = $t_0$ 
from the matrix, $S_f(t,t_0)$, for neutrino flavor f. See Ref\cite{jo04} for a 
detailed derivation of $S_f(t,t_0)$.

  Using the notation $S_{ab}$ and $\bar{S}_{ab}$ for the flavor a,b matrix
element for neutrinos and antineutrinos, the CPV probability is given by 
\beq
\label{CPV}
  \Delta\mathcal{P}^{CP}_{\mu e} &=& \mathcal{P}(\nu_\mu \rightarrow \nu_e)
-\mathcal{P}(\bar{\nu}_\mu \rightarrow \bar{\nu}_e) \nonumber \\
         &=&  |S_{12}|^2- |\bar{S}_{12}|^2 \\
        S_{12} &=& c_{23} \beta -is_{23} a e^{-i\delta_{CP}} A \nonumber \\
  \bar{S}_{12} &=&  c_{23} \bar{\beta} -is_{23} a e^{i\delta_{CP}} \bar{A}
\nonumber \; .
\eeq
The parameters in 
Eq(\ref{CPV}) are
\beq
     a &=&  s_{13}(\Delta -s_{12}^2 \delta) \\
     \delta &=& \delta m_{12}^2/(2 E) \\
      \Delta &=&  \delta m_{13}^2/(2 E) \\
     A & \simeq & f(t) I_\alpha* \\ 
       I_\alpha* &=& \int_0^t dt' \alpha^*(t')f(t') \\
     \alpha(t) &=& cos\omega t -i sin 2\theta sin \omega t \\
          f(t) &=& e^{-i \bar{\Delta} t} \\
 2 \omega &=& \sqrt{\delta^2 + V^2 -2 \delta V cos(2 \theta_{12})} \\
         \beta &=& -i sin2\theta sin\omega L \\
         \bar{\Delta} &=& \Delta-(V+\delta)/2 \\
            sin 2\theta&=& s_{12} c_{12} \frac{\delta}{\omega}  \; ,
\eeq
where the neutrino mass differences are $\delta m_{12}^2=7.6 
\times 10^{-5}(eV)^2$ and $\delta m_{13}^2 = 2.4\times 10^{-3} (eV)^2$.
The neutrino-matter potential $V = \sqrt{2} G_F n_e$, where $G_F$ is the 
universal weak interaction Fermi constant, and $n_e$ is the density of 
electrons in matter. Using the matter density $\rho$=3 gm/cc, which is 
the expected average density for LBNE experiment, $V=1.13 \times 10^{-13}$ eV. 
We use $s_{12} =0.56$ and $s_{23}=0.707$; and $s_{13}=0.15$, as recently found 
in the anti-neutrino disappearance Daya Bay\cite{DB3-7-12} and 
RENO\cite{RENO} experiments. Note that for antineutrinos $\delta_{CP} \rightarrow -\delta_{CP}$.
$\bar{\beta}= \beta (V \rightarrow -V)$ and $\bar{A}= A(V \rightarrow -V)$.

\newpage
For example,

\noindent
2 $\bar{\omega}= \sqrt{\delta^2 + V^2 +2 \delta V 
cos(2 \theta_{12})}$ and $\bar{\bar{\Delta}}=\Delta+(V-\delta)/2$. 

Using conservation of probabiltiy\cite{jo04}, $|A|^2=|\bar{A}|^2$, from 
Eq(\ref{CPV})

\beq
\label{DCPV}
  \Delta\mathcal{P}^{CP}_{\mu e} &=& c_{23}^2(|\beta|^2-|\bar{\beta}|^2)
-2 c_{23} s_{23} a Im[\beta e^{-i\delta_{CP}}A^*- e^{i\delta_{CP}} 
\bar{\beta} \bar{A}^*] \; .
\eeq

 From Eqs(\ref{CPV}-\ref{DCPV}) it follows that 
$\Delta\mathcal{P}^{CP}_{\mu e}$
as a function of energy E and $\delta_{CP}$ is
\beq
\label{DCP(E,CP)}
 \Delta\mathcal{P}^{CP}_{\mu e}(E,\delta_{CP}) &=& c_{23}^2 s_{12}^2 c_{12}^2 
\delta^2
(\frac{s^2}{\omega^2}-\frac{\bar{s}^2}{\bar{\omega}^2}) +2 c_{23}s_{23}
s_{12}c_{12}s_{13}\delta (\Delta-\delta s_{12}^2) \\ 
&&(sin \delta_{CP}
(\frac{s}{\omega}(c-cos\bar{\Delta}L)\frac{\bar{\Delta}-\omega cos 2\theta}
{\bar{\Delta}^2-\omega^2}+\frac{\bar{s}}{\bar{\omega}}(\bar{c}-
cos\bar{\bar{\Delta}}L) \frac{\bar{\bar{\Delta}}-\bar{\omega} cos 2\bar{\theta}}
{\bar{\bar{\Delta}}^2-\bar{\omega}^2}) \nonumber \; \\
&&+cos \delta_{CP} (\frac{s}{\omega (\bar{\Delta}^2-\omega^2))}(sin\bar{\Delta}L
(\bar{\Delta}-\omega cos(2\theta))+
s(\bar{\Delta}cos(2\theta)-\omega)) \nonumber \\
&& -\frac{\bar{s}}{\bar{\omega}(\bar{\bar{\Delta}}^2-\bar{\omega}^2)) }
(sin\bar{\bar{\Delta}}L (\bar{\bar{\Delta}}-
\bar{\omega} cos(2\bar{\theta}))+
\bar{s}(\bar{\bar{\Delta}}cos(2\bar{\theta})-\bar{\omega}))))
 \nonumber \; .
\eeq

  As is clear from Eq(\ref{DCP(E,CP)}), $ \Delta\mathcal{P}^{CP}_{\mu e}
(E,\delta_{CP})$ depends on the value of $s_{13}$ as well as the known
$s_{12},s_{23}$. Fortunately, as shown in Ref\cite{khj11} 
$ \mathcal{P}(\nu_\mu \rightarrow \nu_e)$ and the anti-electron neutrino
dissappearance being studied at Daya Bay, RENO, and Double Chooz is
almost independent of $\delta_{CP}$, and we can use the value $s_{13}=.15$,
consistent with Refs\cite{DB3-7-12,RENO}.

 We calculate $\Delta\mathcal{P}^{CP}_{\mu e}(E,\delta_{CP})$ for L=1300km,
the expected baseline in the proposed LBNE project\cite{LBNE11}. From the
possible energy range E=0.5-12 GeV for the LBNE project\cite{LBNE11} we
estimate $\Delta\mathcal{P}^{CP}_{\mu e}(E,\delta_{CP})$ a function of 
$\delta_{CP}$ for energies within the expected range.

The dependence of on $\Delta\mathcal{P}^{CP}_{\mu e}(E,\delta_{CP})$ on
$\delta_{CP}$ for L=1300km is estimated for LBNE energies E=1, 2, 3, 5, 10 
GeV, as shown in Fig. 1.

\section{Conclusions} 

In the LBNE Report\cite{LBNE07} results of extensive studies have shown that 
future experiments can extend our knowledge of neutrino oscillations beyond 
present and planned experiments. Since there will be both $\nu_\mu$ and 
$\bar{\nu}_\mu$ beams, the LBNE Project can test CPV. 

We have estimated CP violation for the LBNE Project, with a baseline L=1300 km
as a function of $\delta_{CP}$ for $\delta_{CP}$ = 0 to $\pi/2$ for energies
of 1, 2, 3, 5, and 10 GeV, as shown in Fig. 1. We found CPV over 3\% 
with $\delta_{CP}=\pi/2$ for some  energies, which the Project should be 
able to measure. For higher energies, E=5, 10 GeV, 
$\Delta\mathcal{P}^{CP}_{\mu e}$ is smaller than at lower energies, and would
be harder to measure. We find that even for
$\delta_{CP}=0$, for which CPV is entirely a matter effect, CPV probabilities 
of over 1\%  were found for E= 1 GeV, so the Project should be able
to measure CPV for any expected values of $\delta_{CP}$. Fortunately, the
value of $\theta_{13}$ has been determined, which should enable the LBNE
project attain the goal of measuring $\delta_{CP}$.

We believe that our calculations should be useful in planning the future
LBNE Project.

\vspace{3mm}

\Large{{\bf Acknowledgements}}\\
\normalsize
This work was supported in part by a grant from the Pittsburgh Foundation.
The research was carried out in part while LSK was a visitor at LANL P-25, 
and he thanks members of LANL P-25 group for discussions of the LBNE Project. 
Discussions with Ernest Henley and Mikkel Johnson on CPV were very helpful.

\clearpage

\begin{figure}[ht]
\begin{center}
\epsfig{file=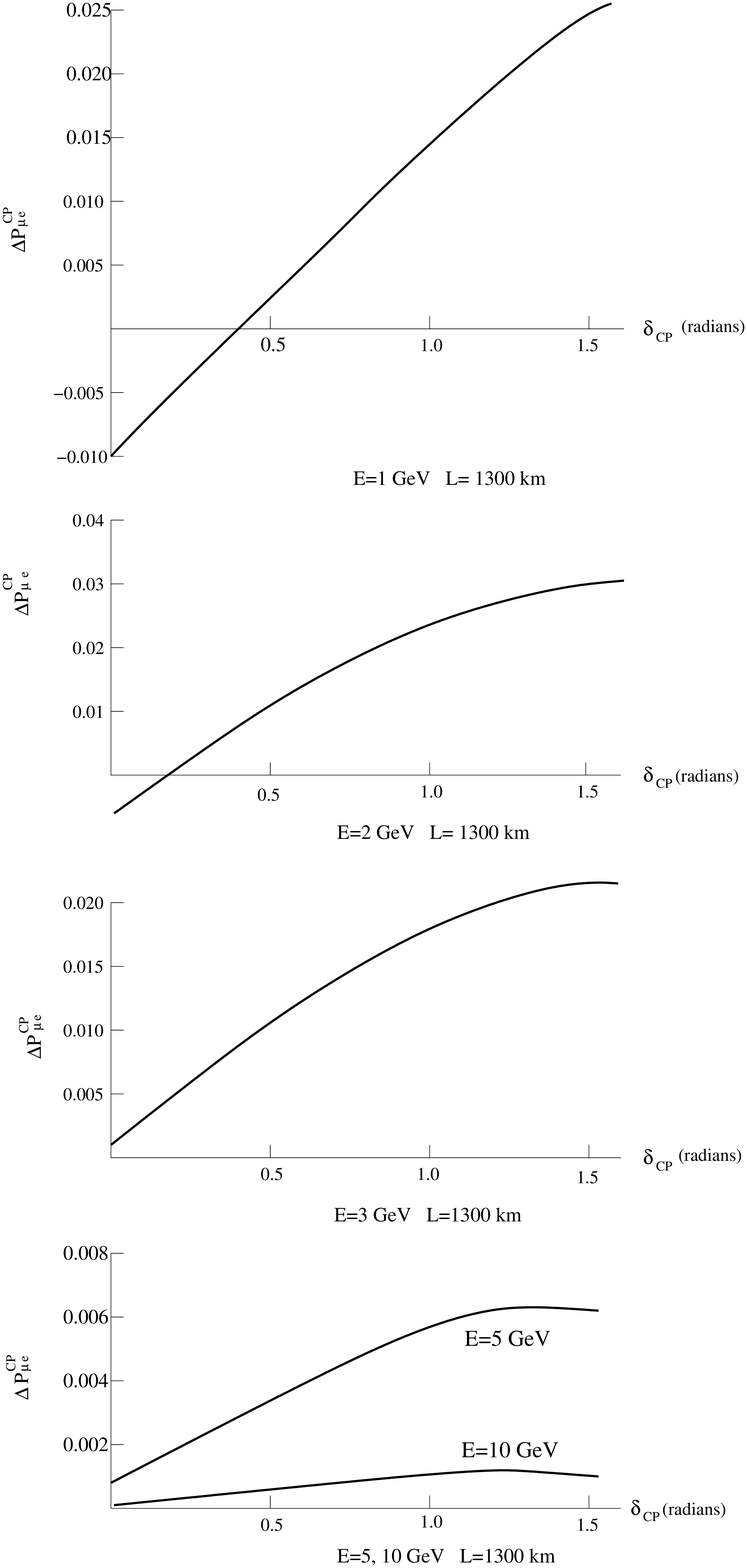,height=21cm,width=12cm}
\end{center}
\caption{$\Delta \mathcal{P}(\nu_\mu \rightarrow\nu_e)$
as a function of $\delta_{CP}$ for expected LBNE L and E}
\end{figure}

\end{document}